\documentstyle[11pt,paspconf]{article}

\input psfig

\begin{document}

\title{The K-band Hubble diagram for the brightest cluster galaxies: 
a  test for galaxy formation and evolution models}

\author{A. Arag\'on-Salamanca$^1$, C.M. Baugh$^2$ \& G. Kauffmann$^3$}
\affil{
$^1$Institute of Astronomy, Madingley Road, Cambridge, England\\
$^2$Department of Physics, South Road, Durham, England\\ 
$^3$Max-Plank-Institut f\"ur Astrophysik, Garching bei M\"unchen, Germany}

\begin{abstract}
The $K$-band Hubble diagram for a sample of brightest cluster galaxies
(BCGs) in the redshift range $0<z<1$ shows a very small scatter
($0.3\,$magnitudes $r.m.s$).  The BCGs exhibit very little luminosity
evolution in this redshift range:  if $q_0=0.0$ we detect {\it no\/}
luminosity evolution; for $q_0=0.5$ we measure a small {\it negative\/}
evolution.  If the mass in stars of these galaxies had remained
constant over this period of time, substantial positive luminosity
evolution would be expected:  BCGs should have been {\it brighter\/} in
the past since their stars were younger. This suggests that the stellar
mass of the BCGs has been assembled over time through merging and
accretion.   
We estimate that the stellar mass in a typical BCG
has grown by a factor $\simeq 2$ since $z\simeq1$ if $q_0=0.0$ or by
factor $\simeq4$ if $q_0=0.5$. These results are in remarkably good
agreement with the predictions of semi-analytic models of galaxy
formation and evolution set in the context of a hierarchical scenario
for structure formation.
\end{abstract}


\section{Introduction}

Brightest cluster galaxies (BCGs) have been extensively studied at
optical wavelengths (see, e.g., Lauer \& Postman 1992; Postman \& Lauer
1995 and references therein). The small scatter in their absolute
magnitudes and their high luminosities make them useful standard
candles for classical cosmological tests involving the Hubble diagram,
such as the determination of $q_0$ and  the variation of $H_0$ with
redshift.  However, there is now firm evidence that significant
evolution has taken place in the colours and optical luminosities of
cluster early-type galaxies (including the BCGs) since $z\simeq1$. 
Thus the Hubble diagram could be seriously affected by
evolutionary changes.  The study of the BCGs in the near-infrared
$K$-band ($2.2\mu$)  has two main advantages over optical studies:
first, the $k$-corrections are appreciably smaller (indeed, they are
negative) and virtually independent of the spectral type of the
galaxies; and second, the light at long wavelengths is dominated by
long-lived stars (see, e.g., Arag\'on-Salamanca et al. 1993). Thus the
$K$-band luminosity is a good measure of the total stellar mass in the
galaxies. In this paper we present new results on the $K$-band Hubble
diagram for the BCGs and compare them with the predictions of
semianalytic models of galaxy formation and evolution. A full account of
these results is given in Arag\'on-Salamanca, Baugh \& Kauffmann (1997).

\begin{figure}

\centerline{\psfig{figure=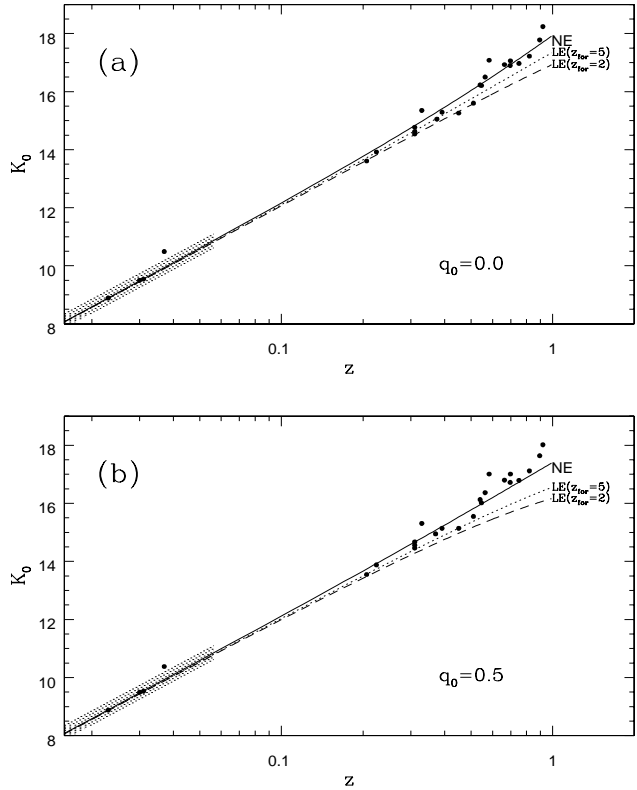,width=90mm,height=115mm}}

\vskip-0.7cm

\caption{{\bf (a)} Magnitude-redshift relation
(Hubble diagram) for the Brightest Cluster Galaxies in the rest-frame
$K$-band.   The solid line shows the no-evolution
prediction (which only takes into account the distance modulus) 
normalised to the low-redshift data. The dashed line
corresponds to the luminosity evolution expected for a stellar
population formed at $z_{\rm for} = 2$ evolving passively.  The dotted
line shows the evolution expected for $z_{\rm for} = 5$.  The shaded
area at low-redshift represents the  $K$-band magnitudes of $z<0.06$
BCGs estimated from the $R$-band data of Postman \& Lauer.
{\bf (b)} As (a) but for $q_0=0.5$.  \label{fig1} }

\section{Observational results}

We analyse a sample BCGs in optically-selected clusters ($0<z<1$) which
should represent the richest clusters present at each redshift. The
$K$-band data come from Arag\'on-Salamanca et al. (1993), Barger et
al.  (1996) and Barger (1997).  Morphological information is available
from ground-based and HST images for many of the clusters.  In general,
the BCGs are either cD galaxies or giant ellipticals.  Photometry for
the BCGs was obtained inside a fixed metric aperture of 50$\,$kpc
diameter ($H_0 = 50\,$km$\,$s$^{-1}\,$Mpc$^{-1}$ assumed
throughout).  Galactic reddening corrections, seeing corrections and
$k$-corrections were applied as in Arag\'on-Salamanca et al. (1993).

\end{figure}

\begin{figure}

\centerline{\psfig{figure=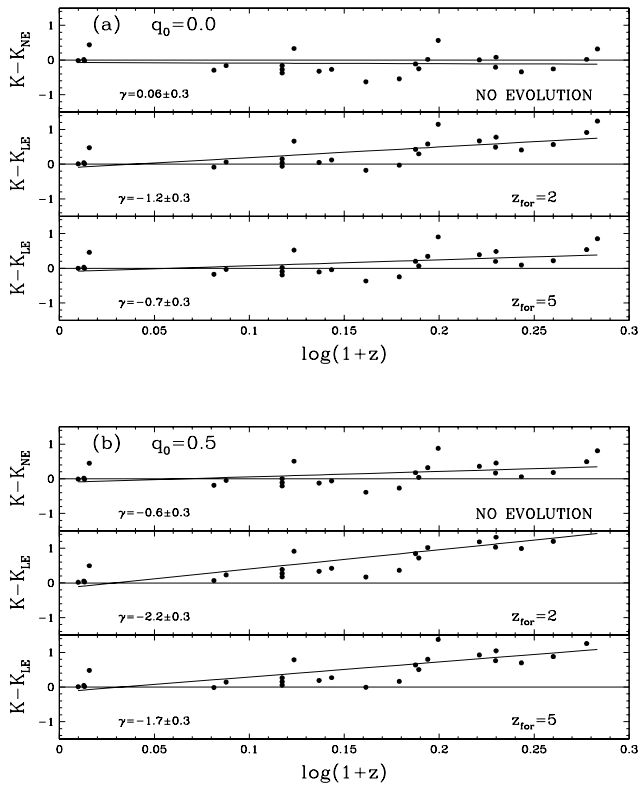,width=90mm,height=115mm}}

\vskip-0.7cm

\caption{{\bf (a)} Top panel: The same data presented in
figure~\ref{fig1} after subtracting the no-evolution prediction. Middle
and bottom panels:  the same data after subtracting models for
luminosity evolution in which the BCG stars form at $z_{\rm for} = 2$
and $z_{\rm for} = 5$ respectively and evolve passively thereafter (for
$q_0=0.0$). The solid lines represent least-squares linear fits of
slope $\gamma$.  {\bf (b)} As (a) but for $q_0=0.5$. \label{fig2} }

\vskip 0.8cm

Figure~\ref{fig1} shows the $K$ magnitude--redshift relation (Hubble
diagram) for the BCGs in our sample. The scatter of the observed
magnitudes around the no-evolution prediction is $0.30\,$mag.  Assuming
that the $K$-band light provides
an estimate of the total stellar mass
of the galaxies, this implies that BCGs at a given
redshift have very similar masses in stars (within 30\% {\it r.m.s.}).
Moreover, the luminosity of the BCGs  does not evolve strongly with
redshift. This is shown more clearly on the top panel of
Figure~\ref{fig2}, where the no-evolution prediction has been
subtracted from the data. For $q_0=0.0$ the observed
magnitudes do not seem to evolve with redshift.  For $q_0=0.5$ there is
a hint of {\it negative\/} evolution:  BCGs at high redshift tend to be
marginally fainter than low redshift ones.  If we parameterise the
luminosity evolution as $L_K(z)=L_K(0)\times(1+z)^\gamma$,
we get $\gamma =-0.06\pm0.3$ and
$-0.6\pm0.3$, for $q_0 = 0.0$ and $0.5$ respectively. Thus the BCGs
show {\it no\/} or marginally {\it negative\/} luminosity evolution.

\end{figure}

However the colours of these galaxies show the same evolution as the
early-type cluster galaxies: they get progressively bluer with redshift
at a rate which indicates that their stellar populations were formed at
$z>2$ and evolved passively thereafter.  If the total stellar mass of
the galaxies has remained constant, we would expect them to get
progressively {\it brighter\/} with redshift, as the average ages of
their stellar populations get younger. Since that brightening is not
observed, the most likely explanation is that the total mass in stars
in the BCGs has grown with time. 
We will now estimate the rate of this growth. 

Using evolutionary population synthesis models (Bruzual \& Charlot
1997) with a standard IMF we predict the expected luminosity evolution
of a passively-evolving stellar population formed at a given redshift.
That should represent the brightening of the stellar populations due to
the decrease in average stellar age with redshift.  The model
predictions are plotted on Figure~\ref{fig1} (LE lines) for formation
redshifts $z_{\rm for} = 2$ and  $z_{\rm for} = 5$.  These models
produce a colour evolution compatible with the observations of
early-type cluster galaxies.  In Figure~\ref{fig2} (lower two panels)
we have plotted the observed magnitudes after subtracting the
luminosity evolution predictions.  Since the models assume a constant
stellar mass, the rate of change with redshift in the $K$ magnitudes,
after correcting for the expected brightening due to passive evolution,
should be a direct measurement of the rate of change in stellar mass.
Parameterising the evolution as
$M_{\star}(z)=M_{\star}(0)\times(1+z)^\gamma$, we obtain the values of
$\gamma$ shown in Figure~\ref{fig2}, which imply that the mass in stars
of a typical BCG grew by a factor $\simeq2$ if $q_0 = 0.0$ or $\simeq4$
if $q_0 = 0.5$ from $z\simeq1$ to $z\simeq0$.

\section{Galaxy formation and evolution models}

The recent development of {\it semi-analytic techniques\/} has provided
theorists with the tools to make predictions for the formation and
evolution of galaxies, using physically motivated models set in the
context of hierarchical structure formation in the universe.  The
properties of galaxies in these models are in broad agreement with the
present day characteristics of galaxies, such as the distribution of
luminosities, colours and morphologies and with the properties of
galaxies at high redshift, including the faint galaxy counts, colours
and redshift distributions. Full details of the models and these
results can be found in Kauffmann et al. (1993, 1994) for the  Munich
models and Cole et al. (1994) and Baugh et al. (1996, 1997) for the
Durham models. See also the contribution by Frenk in this volume.  In
these models, there are three different ways in which the stellar
masses of the brightest galaxies in clusters can grow:  (1) merging of
satellite galaxies that sink to the center of the halo through
dynamical friction; (2) quiescent star formation as a result of gas
cooling from the surrounding hot halo medium; and (3) ``bursts'' of
star formation associated with the merging of a massive satellite.

\begin{figure}

\centerline{\psfig{figure=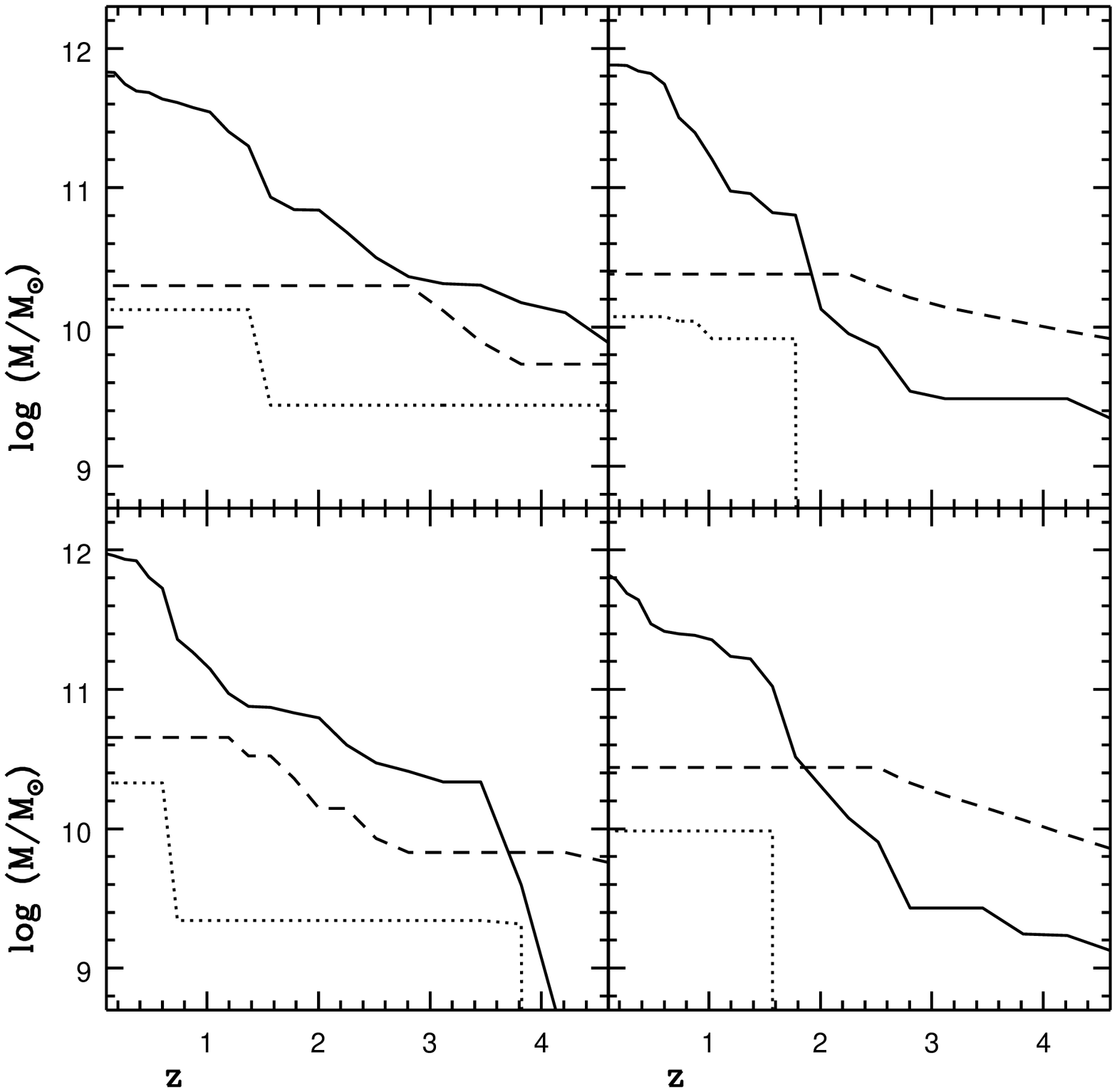,width=60mm}\psfig{figure=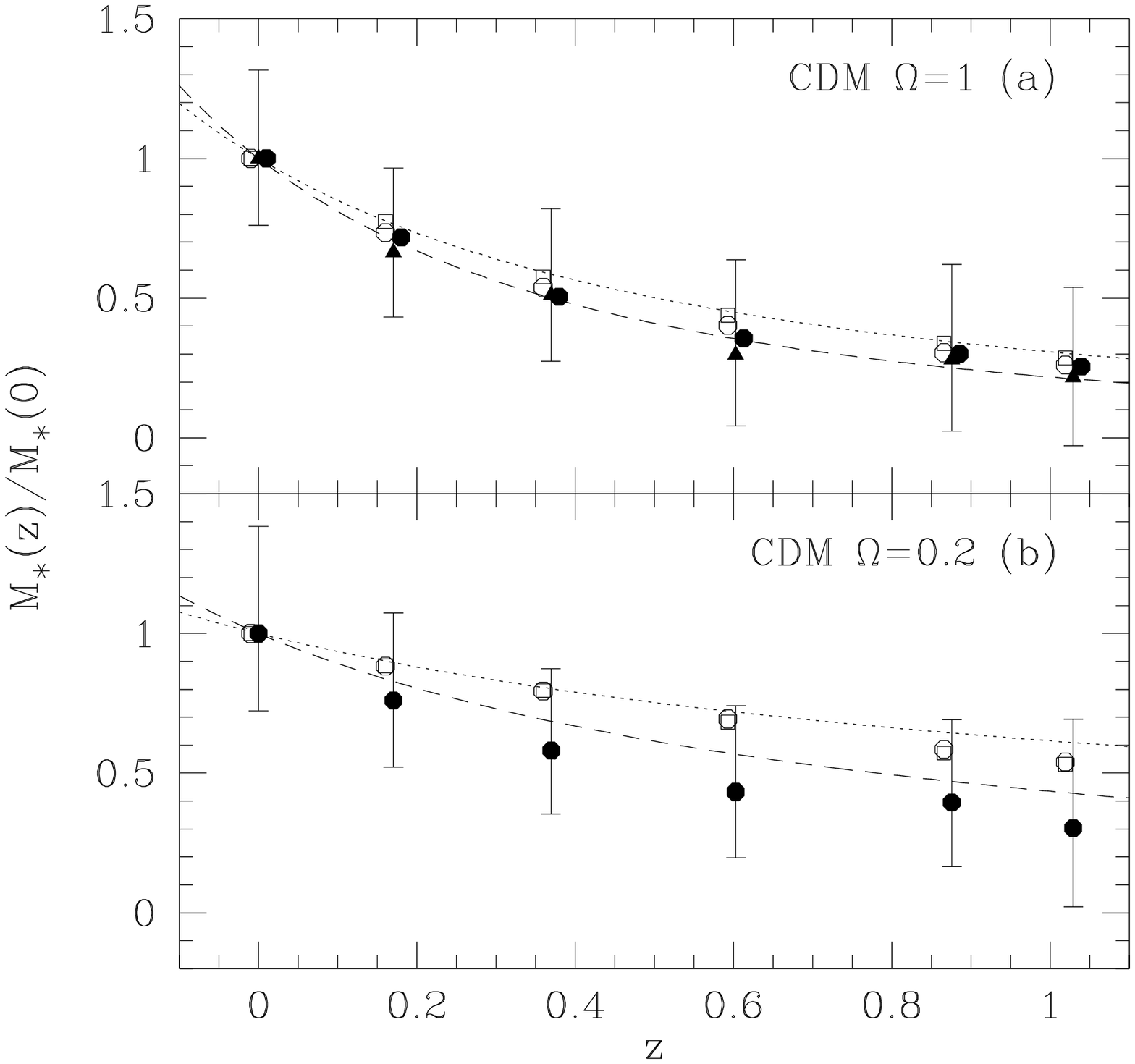,width=63mm}}

\vskip -0.5cm

\caption{{\bf (Left)} The build up in stellar mass of the largest
progenitor in four example BCGs.  The four panels represent independent
Monte Carlo realizations of the formation of a BCG in a halo of
circular velocity $1000\,$km$\,$s$^{-1}$.  The solid lines show the
accumulation of stellar mass from merging events, the dashed lines show
the mass contributed by stars forming from gas cooling from the
surrounding hot halo medium, and the dotted lines show the mass
contributed by star formation bursts associated with major mergers.  A
Cold Dark Matter cosmology with critical density was assumed.
\label{fighist}}

\caption{{\bf (Right)} The average stellar mass of BCGs in halos of
circular velocity $1000\,$km$\,$s$^{-1}$, normalised to the average BCG
mass at redshift zero.  {\bf (a)} Predictions for a Cold Dark
Matter universe with the critical density. {\bf (b)} Predictions for an open
universe.  The density fluctuations in each case are normalised to
reproduce the abundance of rich clusters.  The open symbols show Durham
models for isothermal halos (circles) and for halos with Navarro, Frenk
\& White (1996) density profiles (squares).  The filled symbols show
the Munich models; for the case of the filled circles, no visible stars
form in the cooling flows of halos with circular velocities greater
than $500\,$km$\,$s$^{-1}$. The filled triangles are for a model in
which stars do form in massive cooling flows. The curves show the
trends in stellar mass deduced in Section 2; the dotted curve
corresponds to $z_{for}=5$ and the dashed curve to $z_{for}=2$.  The
error bars shown are representative of the scatter found in all the
models ($\simeq30\,$percent, i.e., very close to the observed one).
\label{figmass}}

\vskip 0.8cm

\noindent too blue and too bright if all gas in the central cooling
flows of massive clusters turns into stars with a normal IMF.  The
Munich models adopt a somewhat {\it ad hoc\/} fix by simply switching
off star formation in cooling flows once the circular velocity of a
halo grows larger than $500\,$km$\,$s$^{-1}$ (this value was chosen so
that the models produced a good fit to the bright end of the Virgo
cluster luminosity function).  Figure~\ref{fighist} shows that star
formation from bursts and cooling gas then only account for a few
percent of the final mass of the BCG. The rest comes from accreted
galaxies.

\end{figure}

\section{Discussion}

In figures~\ref{fighist} and~\ref{figmass} we illustrate how $z=0$
BCGs are predicted to evolve by the semi-analytic models. One
unresolved problem in  these models is that the colours and
absolute magnitudes of central cluster galaxies are predicted to$\,$ be 


The observational selection picks out {\it the richest clusters\/} at
each redshift. We mimic this selection by calculating the masses of
BCGs in halos {\it of fixed circular velocity\/} as a function of
redshift. This means we are selecting rarer objects with increasing
redshift. In  Figure~\ref{figmass} we compare both the Durham and the
Munich models with the observations.  Note that different assumptions
about the dark halo profiles or the formation of stars in cooling flows
make very little difference to the predictions for the {\it relative
change\/} in the masses of BCGs from $z=0$ to $1$. This is because the
crucial parameter controlling this change is the {\it time since the
Big Bang}. High redshift clusters have less massive BCGs because there
has been less time for these galaxies to assemble as a result of
merging or gas accretion.

The agreement between the models and the observations in both the rate
of growth of the stellar mass and the scatter is remarkable:  both
models agree well with the data when the appropriate value of $q_0$ is
used in the analysis. It is thus not possible to make statements about
a preferred cosmology.

\acknowledgments

We thank G.~Bruzual and S.~Charlot for allowing us to use their model
results prior to publication, and A. Barger for some of the $K$-band
data.  The current Durham semi-analytic models are the result of a
collaboration between Shaun Cole, Carlos Frenk, Cedric Lacey and CMB.
This work was carried out under the auspices of EARA, a European
Association for Research in Astronomy, and the TMR Network on Galaxy
Formation and Evolution funded by the European Commission.

\end{document}